\documentclass[twocolumn,trackchanges]{aastex631}

\hypersetup{linkcolor=red,citecolor=blue,filecolor=cyan,urlcolor=magenta}

\begin{document}

\title{Minute time scale variability in $\gamma$-ray flare of BL Lacertae}

\author{Joysankar Majumdar}
\affiliation{Department of Physics, Institute of Science, Banaras Hindu University, Varanasi-221005, India}

\author{Sakshi Maurya}
\affiliation{Department of Physics, Institute of Science, Banaras Hindu University, Varanasi-221005, India}

\author{Raj Prince}
\affiliation{Department of Physics, Institute of Science, Banaras Hindu University, Varanasi-221005, India}
\email{priraj@bhu.ac.in}



\begin{abstract}
In October 2024, The object BL Lac experienced a brightest flaring event in gamma-ray ($>$100 MeV) with a historical $\gamma$-ray flux of $\sim$10$^{-5}$ erg cm$^{-2}$ s$^{-1}$. Soon after the event was followed across the waveband and in X-ray (0.3-10 keV) it was also found to be flaring with the maximum flux achieved during this event as 8.30$\times$10$^{-11}$ erg cm$^{-2}$ s$^{-1}$. The high gamma-ray significance enables us to probe the shortest time scale variability possible and for that, we produced the orbital binned light curve, 5 minutes binned light curve, and the 2 minutes binned light curve. A clear variation is seen in the 5-minute light curve and is fitted with the sum of exponentials to derive the rise and decay time scale which ranges between 3 to 12 minutes. The fastest variability time is also estimated to be an order of 1 minute from 2 minutes. The estimated size of the emission region is very small (10$^{13}$ cm) compared to the size of the black hole event horizon. The location of the emission region is estimated to be very close to the supermassive black hole (10$^{14}$ cm) and much inside the BLR (0.1 pc). We discussed the possible way to explain this fast-flux variability in BL Lac.

\end{abstract}


\section{Introduction} \label{sec:intro}
Blazars are an interesting type of active galactic nuclei (AGNs) that have highly relativistic jets along the line of sight (angles less than $\sim 14^\circ$ ) to the observer \citep{1995PASP..107..803U}. The highly luminous and powerful jets of blazars are powered by massive black holes and they dominate the extra-galactic $\gamma$-ray sky. Blazars are further classified into two groups- flat-spectrum radio quasar (FSRQ) and BL Lac objects (BL Lacs). It is generally assumed that shocks traveling in the jet or the jet formation region near the central black hole may produce the $\gamma$-ray emission from blazars. Minute-scale variability indicates highly compact emission region (e.g. \citealt{2018ApJ...854L..26S}) in blazars. Previously, \citealt{2014Sci...346.1080A} found about 4.8 minutes $\gamma$-ray variability in IC 310 and \citealt{2018ApJ...854L..26S} also found 5 minutes $\gamma$-ray variability in CTA 102.

BL Lacertae is the archetype of BL Lacs with a redshift of $z = 0.0668 \pm 0.0002$ \citep{1977ApJ...212...94H}. On 27 April 2021, the brightest $\gamma$-ray flux ever detected from BL Lacertae. With 2-min binned $\gamma$-ray light-curve, \citealt{2022A&A...668A.152P} reported nearly 1 minute $\gamma$-ray variability, which is the shortest GeV variability timescale ever observed from blazars. In October 2024, Fermi-LAT reported enhanced $\gamma$-ray activity in BL Lacertae (ATel\#16849, \citealt{2024ATel16849....1V}). Following this VERITAS detected $\gamma$-ray flaring (ATel\#16854 \citealt{2024ATel16854....1C}), MAGIC detected very-high-energy $\gamma$-ray (ATel\#16861 \citealt{2024ATel16861....1P}) and high activity in optical band detected in DFOT (Atel\#16856 \citealt{2024ATel16856....1K}) and LAST (ATel\#16865 \citealt{2024TNSAN.294....1G}) on October 2024.

Our motivation in this paper is to find rapid temporal flux variation in the recent flare and to predict the emission region size that is responsible for this ultra-fast $\gamma$-ray variability.

\section{Observations and data analysis} \label{sec:obs_data}
\subsection{Fermi-LAT} \label{subsec:lat}
We have used $\gamma$-ray data from Fermi Large Area Space Telescope (LAT) from 22 August 2024 to 15 October 2024 in 100 MeV to 500 GeV range within 20$^\circ$ radius around the source. We used \texttt{FermiPy}\footnote{\url{https://github.com/fermiPy/fermipy}} \citep{2017ICRC...35..824W} PYTHON package to analyze the data which uses the Fermi Science Tools\footnote{\url{https://fermi.gsfc.nasa.gov/ssc/data/analysis/documentation}} in back-end. \texttt{FermiPy} requires a YAML file where all the analysis and data selection parameters are defined. A circular region of interest (ROI) of size 10$^\circ$ was chosen for the analysis of all the sources in the same period MJD 60544–60598. We constrained the data with \textit{evclass=128} and \textit{evtype=3} for \texttt{gtselect} and applied zenith angle cut-off to be 90$^\circ$. We also used the filter \texttt{DATA\textunderscore QUAL$>$0\&LAT\textunderscore CONFIG==1} for the \textit{gtmktime} tool. We used Galactic interstellar emission model (\texttt{gll\textunderscore iem\textunderscore v07}\footnote{\url{https://fermi.gsfc.nasa.gov/ssc/data/access/lat/BackgroundModels.html}}) and isotropic diffuse emission template (\texttt{iso\textunderscore P8R3\textunderscore SOURCE\textunderscore V3\textunderscore v1}\footnote{\url{https://fermi.gsfc.nasa.gov/ssc/data/access/lat/BackgroundModels.html}}). We kept free the normalization of all sources within 3$^\circ$ of the ROI center and freed all the parameters of isotropic and galactic diffuse components. We performed binned likelihood analysis with 8 energy bins per decade and a spatial bin size of 0.1$^\circ$. While generating spectral energy distribution (SED) we took 4 energy bins per decade in the configuration YAML file.

\begin{figure*}[ht!]
    \gridline{\fig{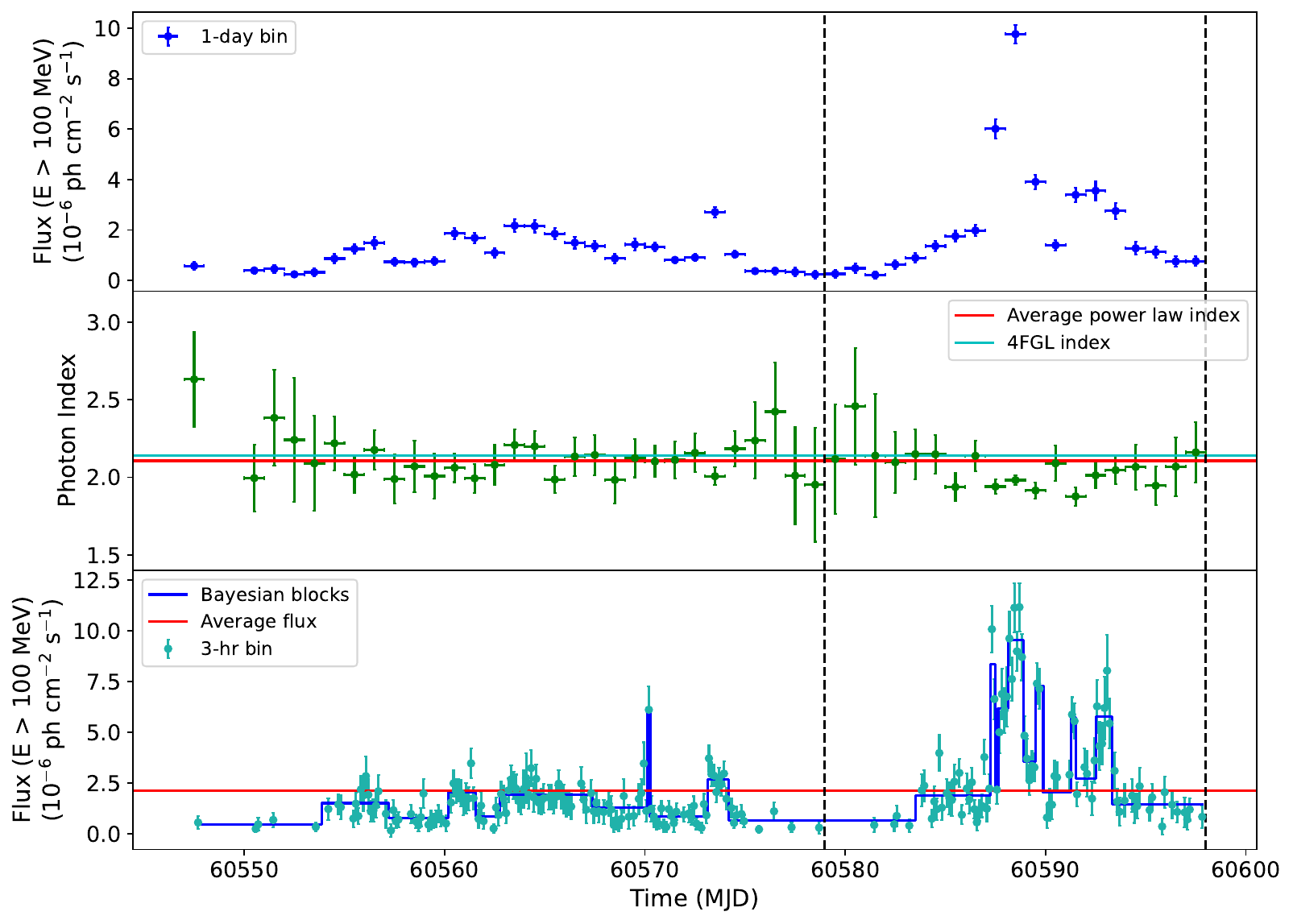}{0.48\textwidth}{(a)}
              \fig{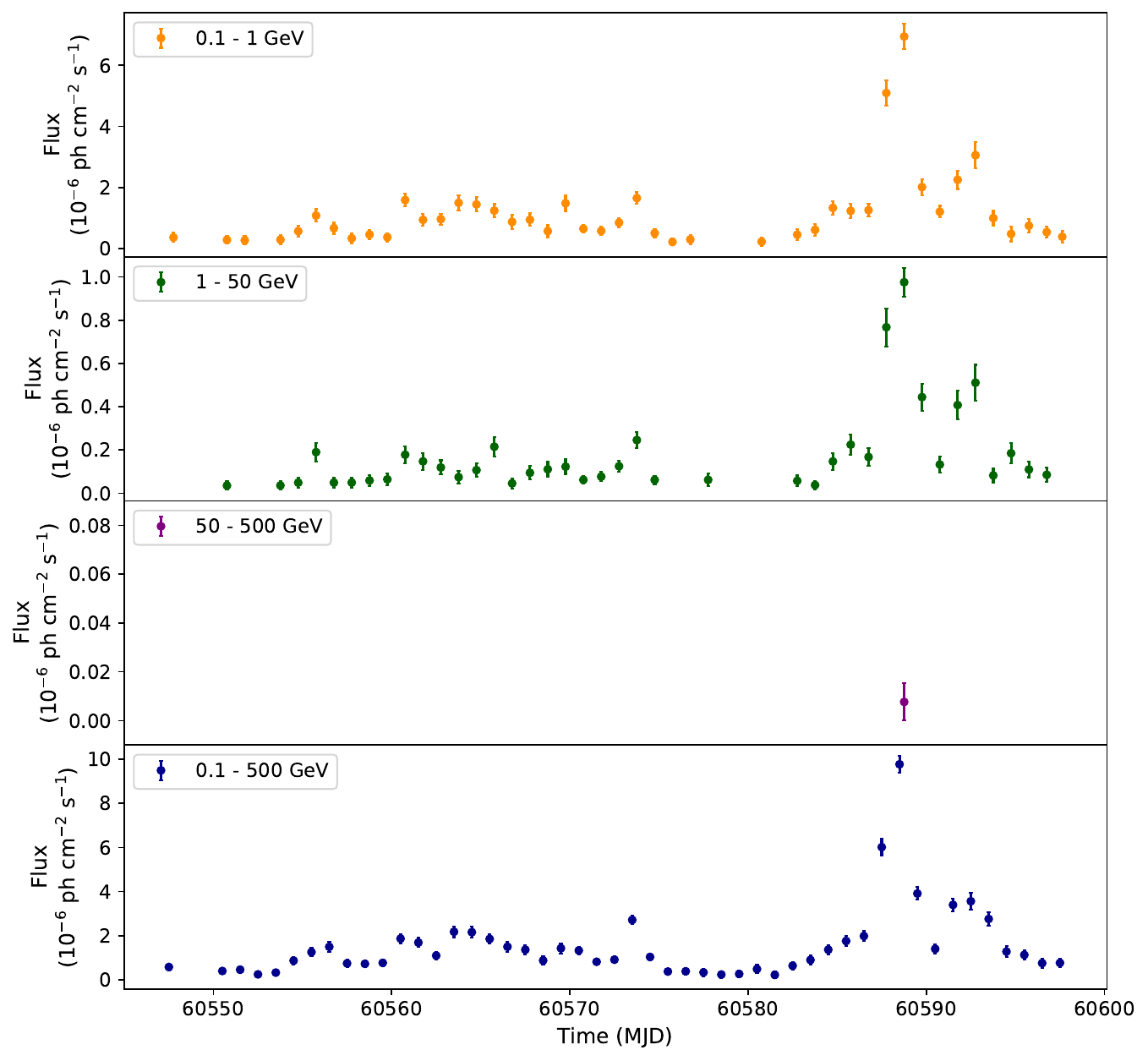}{0.48\textwidth}{(b)}}
\caption{(a) Fermi-LAT light curves of BL Lacertae from MJD 60544 to MJD 60598. The first panel shows a 1-day binned light curve in the 0.1 - 500 GeV energy range. The second panel shows the temporal variation of the $\gamma$-ray photon index. The red line indicates the average PL index and the cyan line indicates the PL index mentioned in the 4FGL catalog. The third panel shows the 3-hr binned light curve. The blue lines indicate the Bayesian blocks and the red line indicates the average flux. (b) 1-day binned $\gamma$-ray light curve in the different energy range of BL Lacertae. The energy range used to generate the light curve is mentioned in each panel.
\label{fig:gamma_lc}}
\end{figure*}

\subsection{Swift-XRT} \label{subsec:xrt}
Data from the X-ray focusing telescope (XRT; 0.3–10.0 keV) of Neil Gehrels Swift Observatory from MJD 60591.72 to MJD 60592.71 is used in this work. We followed the standard data reduction procedure by extracting the source and background counts from the circular region of 10 and 30 arcsec radius around the source and away from the source, respectively. We used \texttt{XSELECT} to generate source and background spectrum and used \texttt{XSPEC} to model the spectrum. We modeled the spectrum with a simple power law (PL) as follows
\begin{equation}
    N(E) = N_0\ E^{-\alpha},
\end{equation}
where N$_0$ is the normalization factor and $\alpha$ is the power law index. The interstellar absorption has been corrected with the H$_{I}$ column density, n$_{H}$ = 2.92 $\times$ 10$^{21}$ cm$^{-2}$. In this way, we fitted every observation and estimated the flux and the corresponding power-law index.

\section{Results} \label{sec:results}
\begin{figure*}
    \gridline{\fig{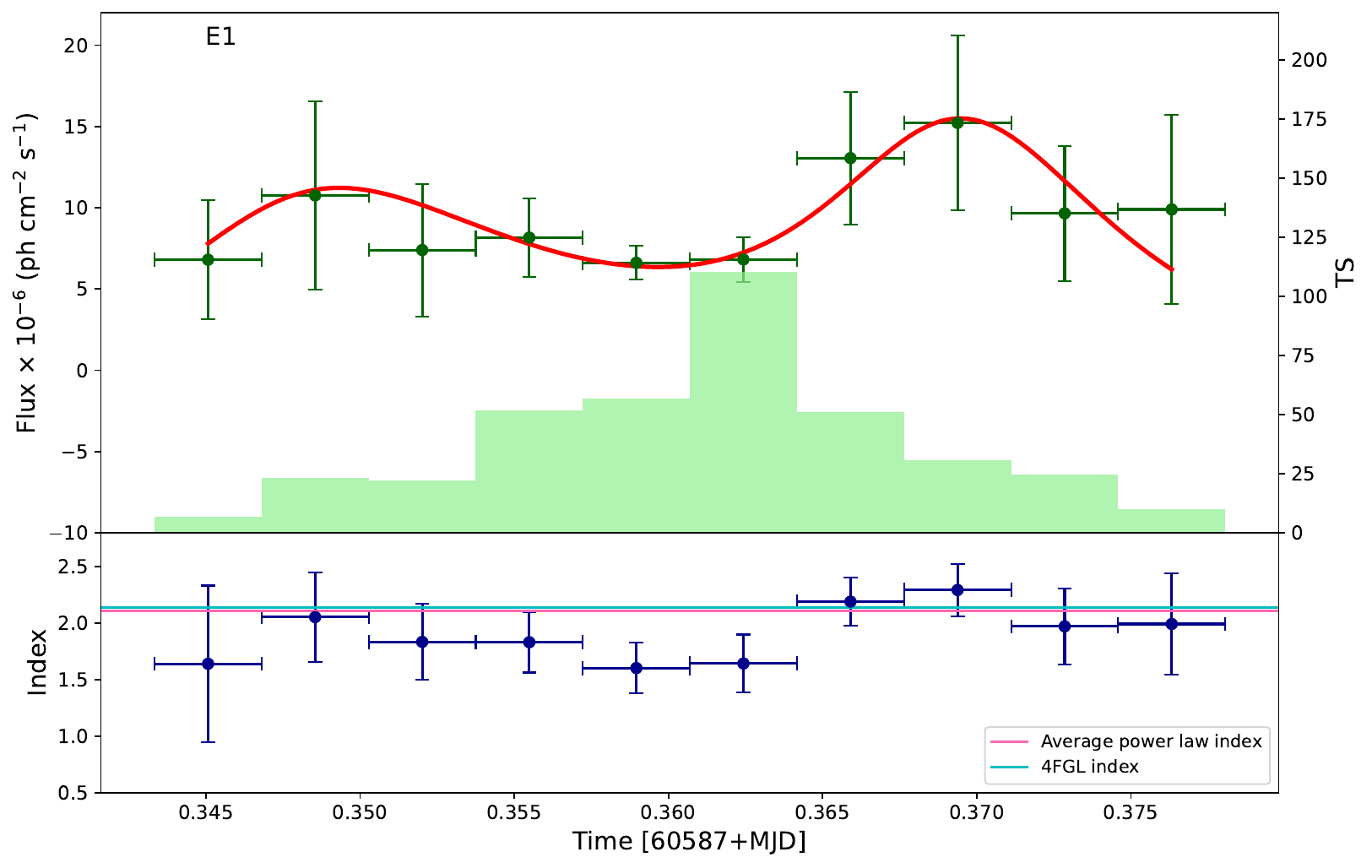}{0.33\textwidth}{(a)}
              \fig{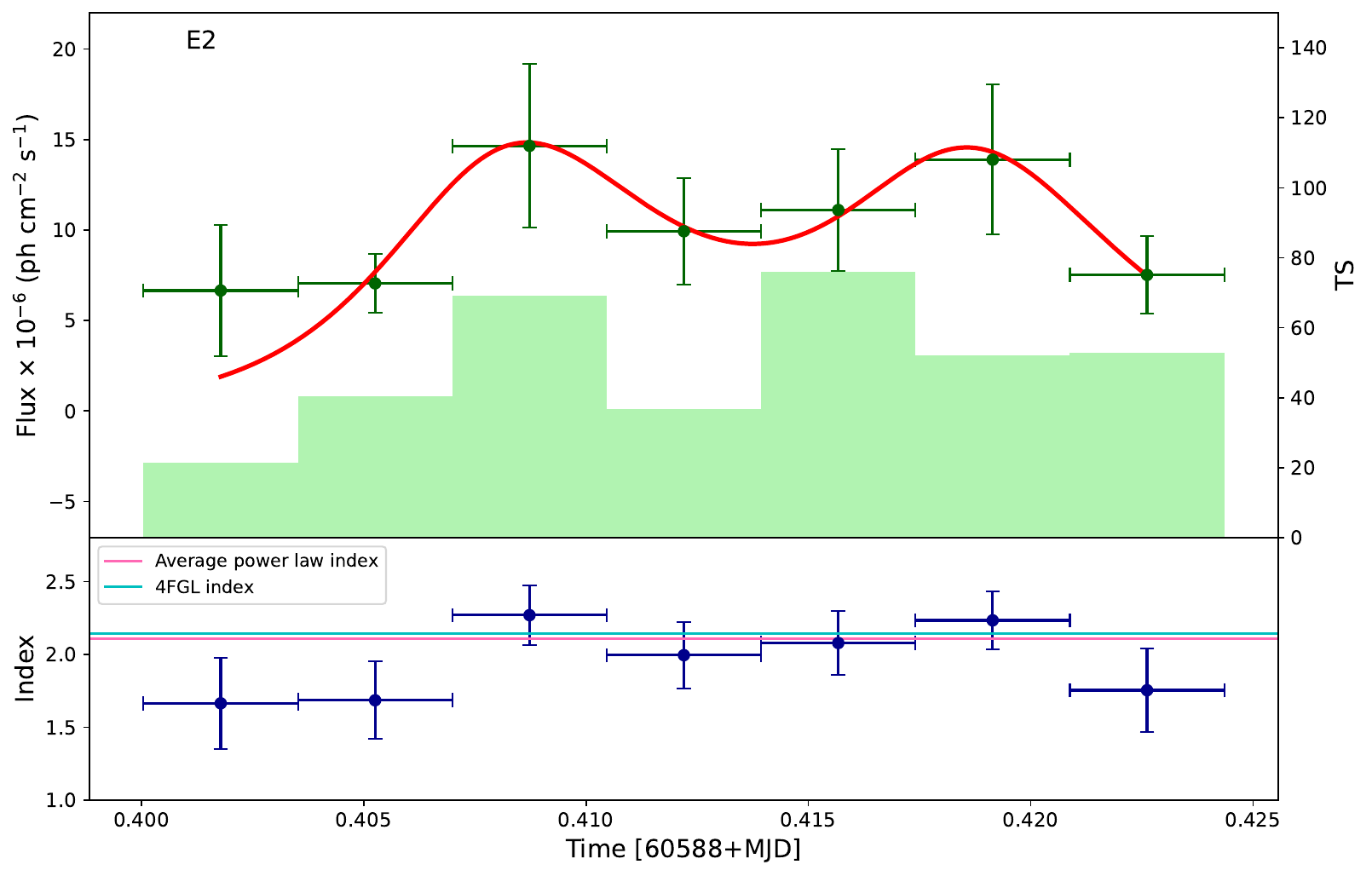}{0.33\textwidth}{(b)}
              \fig{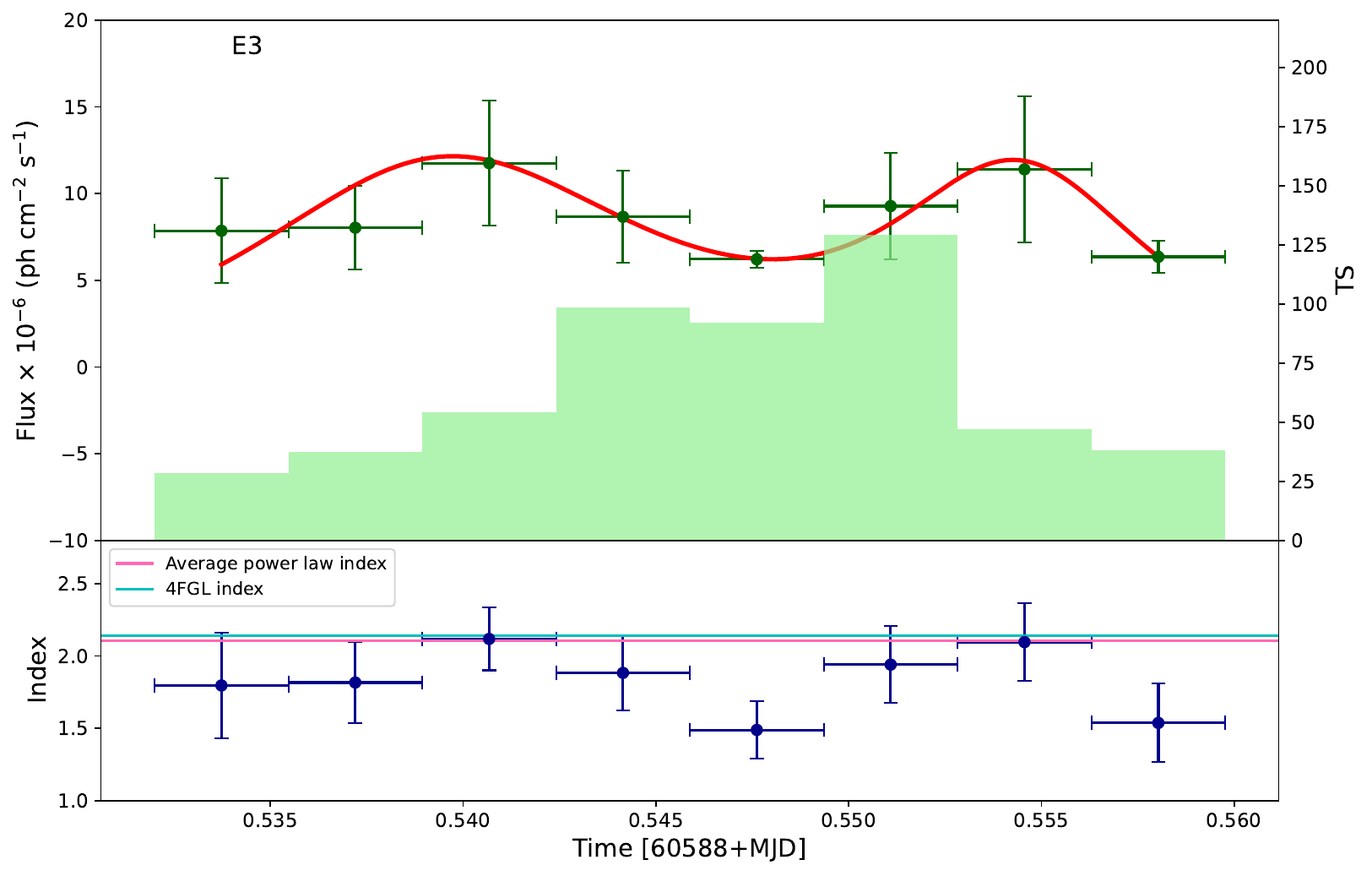}{0.33\textwidth}{(c)}}
    \gridline{\fig{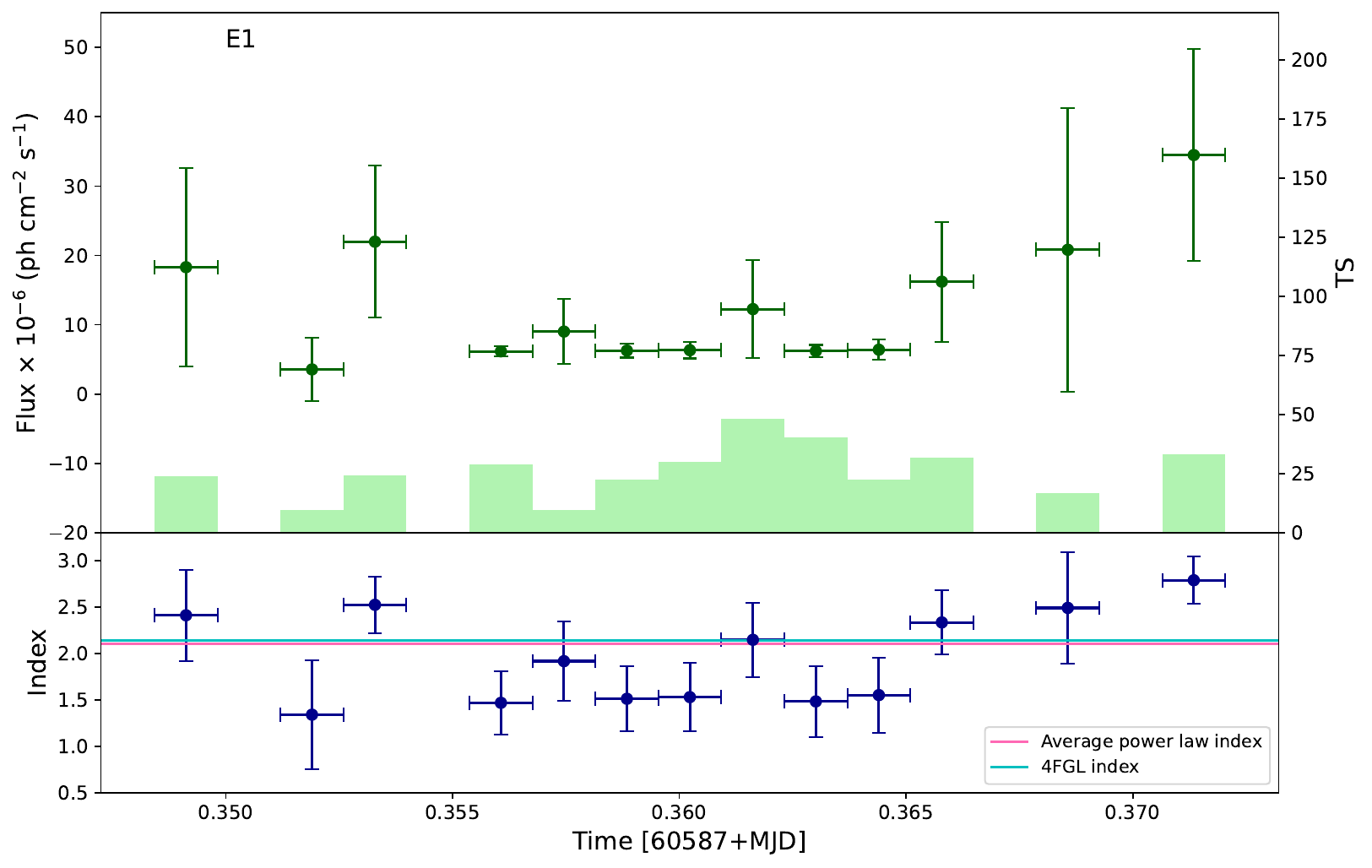}{0.33\textwidth}{(d)}
              \fig{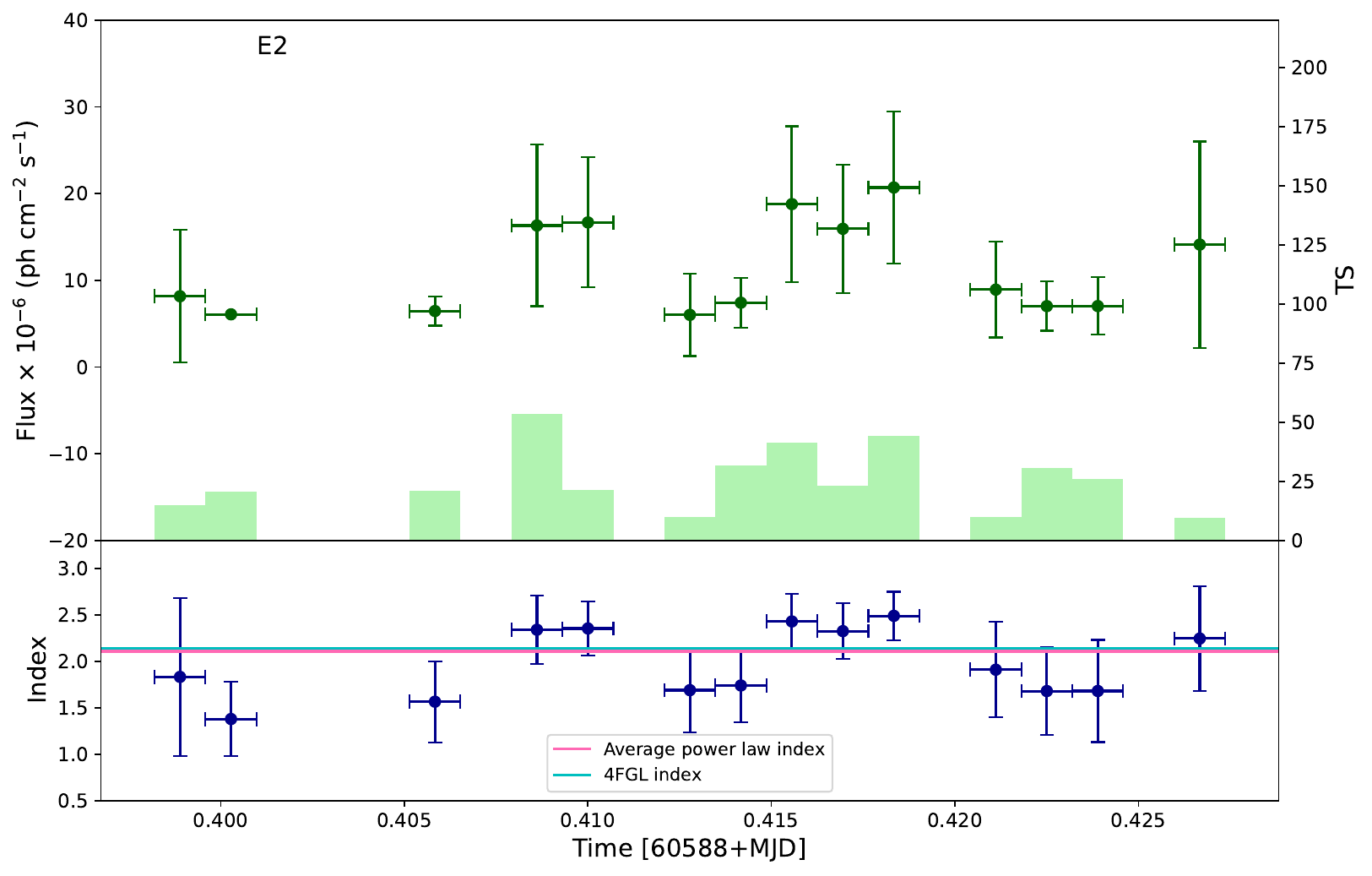}{0.33\textwidth}{(e)}
              \fig{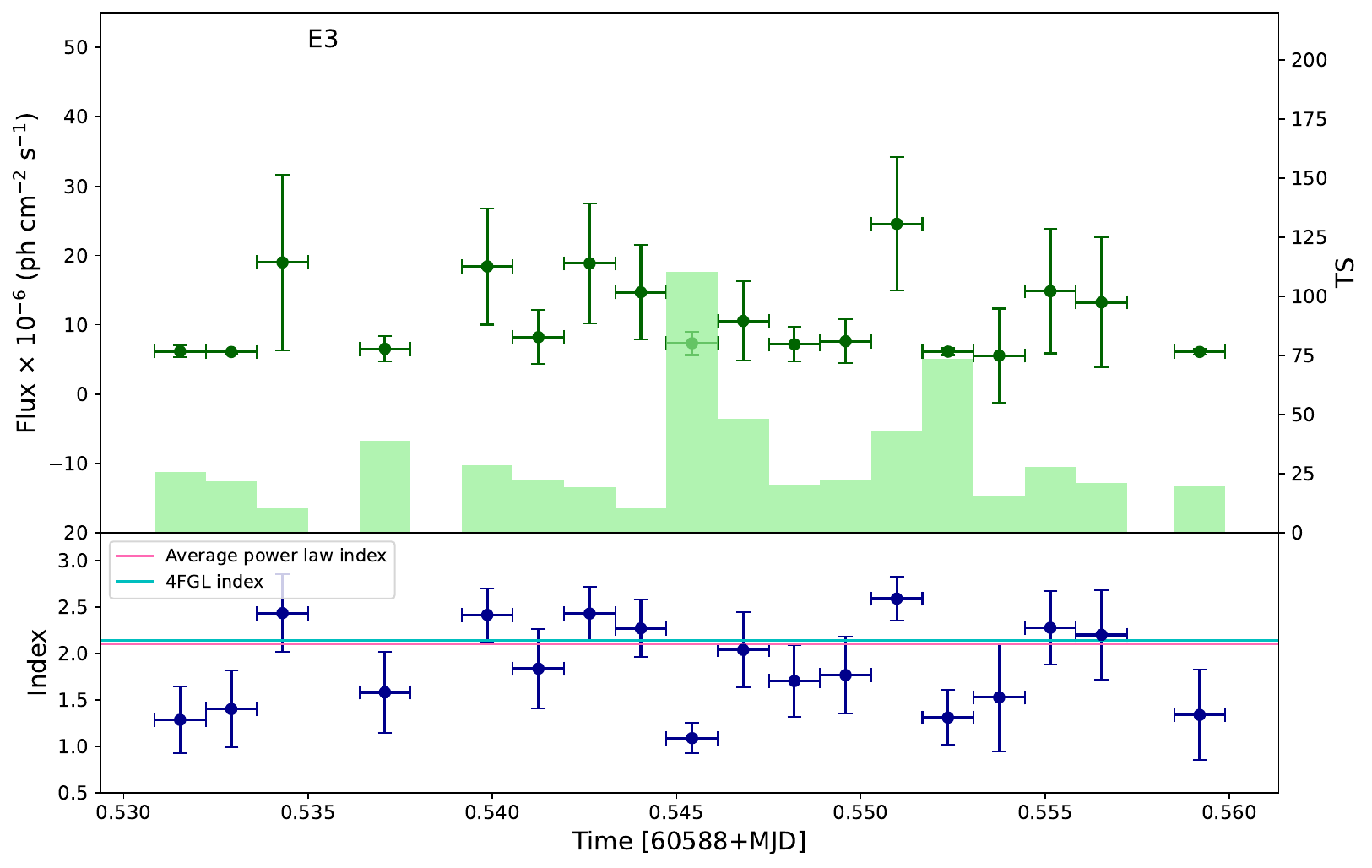}{0.33\textwidth}{(f)}}
    \caption{Upper panel of (a), (b) and (c) show 5min-binned light curves of BL Lacertae in different orbits with TS value in the green bar and the lower panel shows the variation of PL index for that period. The red line is the peak fitted sum of the exponential function and their parameters are mentioned in Table \ref{tab:rise_decay_table}. Whereas (d), (e) and (f) show the same but with 2min-binned data.\label{fig:short_lcs}}
\end{figure*}

\subsection{$\gamma$-ray light curve} \label{subsec:gamma_lc}
The spectral shape of BL Lacertae is defined as log-parabola in Fermi 4FGL catalog \citep{2020ApJS..247...33A} but we generated the light curve using \texttt{FermiPy} by modeling the spectra in each time bin with a power law (PL). The power law model is more suitable for this work as we are dealing with shortest scale $\gamma$-ray flux variation. The 1-day binned $\gamma$-ray light curve has been shown in the first panel of Figure \ref{fig:gamma_lc}(a) from MJD 60544 to MJD 60598. In this particular flaring period of BL Lacertae, we got the historical height daily binned flux (9.77 $\pm$ 0.37) $\times$ 10$^{-6}$ ph cm$^{-2}$ s$^{-1}$ on 5th October 2024. The photon index calculated using PL for this brightest flux is 1.98 $\pm$ 0.03 which is slightly harder than the photon index value (2.143) mentioned in the Fermi 4FGL catalog.

We implemented the Bayesian blocks (BB) algorithm \citep{2013ApJ...764..167S} false-alarm probability (p$_0$) of 0.05 to identify the flare in 3-hr binned $\gamma$-ray light curve with the help of a PYTHON package \texttt{lightcurves}\footnote{\url{https://github.com/swagner-astro/lightcurves}} \citep{2022icrc.confE.868W} shown in the last panel in Figure \ref{fig:gamma_light_curve}. With the help of the HOP algorithm, we identify the flare period with a condition F$_{BB} \geq 3\bar{F}$, mean flux. Thus we have selected MJD 60578.98 to MJD 60598.00 as our main flaring period indicated in the black dotted line Figure \ref{fig:gamma_lc}(a).

We also generated 1-day binned light curve of BL Lacertae in 0.1-1 GeV, 1-50 GeV, and 50-500 GeV energy bands for that flaring period chosen previously with test statistic (TS) $> 9$ following \citealt{2022A&A...668A.152P}. As seen in Figure \ref{fig:gamma_lc}(b), light curves in 0.1-1 GeV and 1-50 GeV energy bands follow a similar trend of 0.1-500 GeV energy band whereas only one point is detected in 50-500 GeV energy band.

\subsection{Minute-scale variability} \label{subsec:minute_scale_var}
We further investigate from MJD 60578.98 to MJD 60598.00 as we can see in the third panel of Figure \ref{fig:gamma_lc}(a), fluxes is around the 1 $\times$ 10$^{-5}$ ph cm$^{-2}$ s$^{-1}$. We generated the light curves of this period with bin size equal to the orbital period ($\sim$ 95 min) using the Fermi-LAT data. From the orbital binning $\gamma$-ray light curve, we have selected two sections, S1 from MJD 60587.28003472 to MJD 60587.54623843 and S2 from MJD 60588.29855324 to MJD 60588.70364583 as they have higher TS value (TS $> 250$) to generate 5 minutes binning $\gamma$-ray light curve. We have two chosen the orbit E1 lies within S1 and two orbits E2 and E3 from S2. We have estimated the shortest flux doubling/halving timescale from these light curves, as follows:
\begin{equation} \label{eq:flux_doubling_time}
    F(t_2) = F(t_1) \times 2^{\frac{t_d}{\Delta t}}
\end{equation}
where F($t_1$) and F($t_2$) are the flux values at times $t_1$ and $t_2$ , respectively, $\Delta t = t_1 - t_2$ , and $t_d$ denotes the flux doubling/halving timescale. We observed the shortest flux halving timescale about 32.8 minutes from the orbital binned light curve.
To study the temporal evaluation of flux, we fitted the peaks of a 5min-binned light curves with the sum of exponential in Figure \ref{fig:short_lcs} (a), (b) and (c) defined as
\begin{equation} \label{double_exp}
F(t) = 2F_0 \left[\exp\left(\frac{t_0 - t}{T_R}\right) + \exp\left(\frac{t - t_0}{T_D}\right)\right]^{-1} ,
\end{equation}
where $T_R$ and $T_D$ are the rise and decay times of a particular flare and $F_0$ is the peak flux observed at time $t_0$. The shortest flux doubling/halving time, rise time, and decay time for each peak is defined in Table \ref{tab:rise_decay_table}. We searched for minute-scale temporal variability of flux further by generating a 2 min-binned light curve of this period as we can see rapid flux variations with high TS values (e.g. \citealt{2018ApJ...854L..26S,2022A&A...668A.152P}). The 2min-binned light curves are shown in Figure \ref{fig:short_lcs} (d), (e) and (f). We have also calculated the minimum flux doubling/halving timescale using equation \ref{eq:flux_doubling_time} within the 2min-binned light curve. For E1, we found $t_d$ $\approx$ 1.49 min where flux changed from (6.40 $\pm$ 1.48) $\times$ 10$^{-6}$ to (1.62 $\pm$ 0.86) $\times$ 10$^{-5}$ ph cm$^{-2}$ s$^{-1}$. For E2, we found $t_d$ $\approx$ 1 min where flux changed from (2.46 $\pm$ 0.96) $\times$ 10$^{-5}$ to (6.12 $\pm$ 0.49) $\times$ 10$^{-6}$ ph cm$^{-2}$ s$^{-1}$.And for E3, we found $t_d$ $\approx$ 1.48 min where flux changed from (7.40 $\pm$ 2.85) $\times$ 10$^{-6}$ to (1.88 $\pm$ 0.89) $\times$ 10$^{-5}$ ph cm$^{-2}$ s$^{-1}$.

\subsection{$\gamma$-ray spectral variability} \label{subsec:gamma_spec_var}
\begin{figure*}
    \gridline{\fig{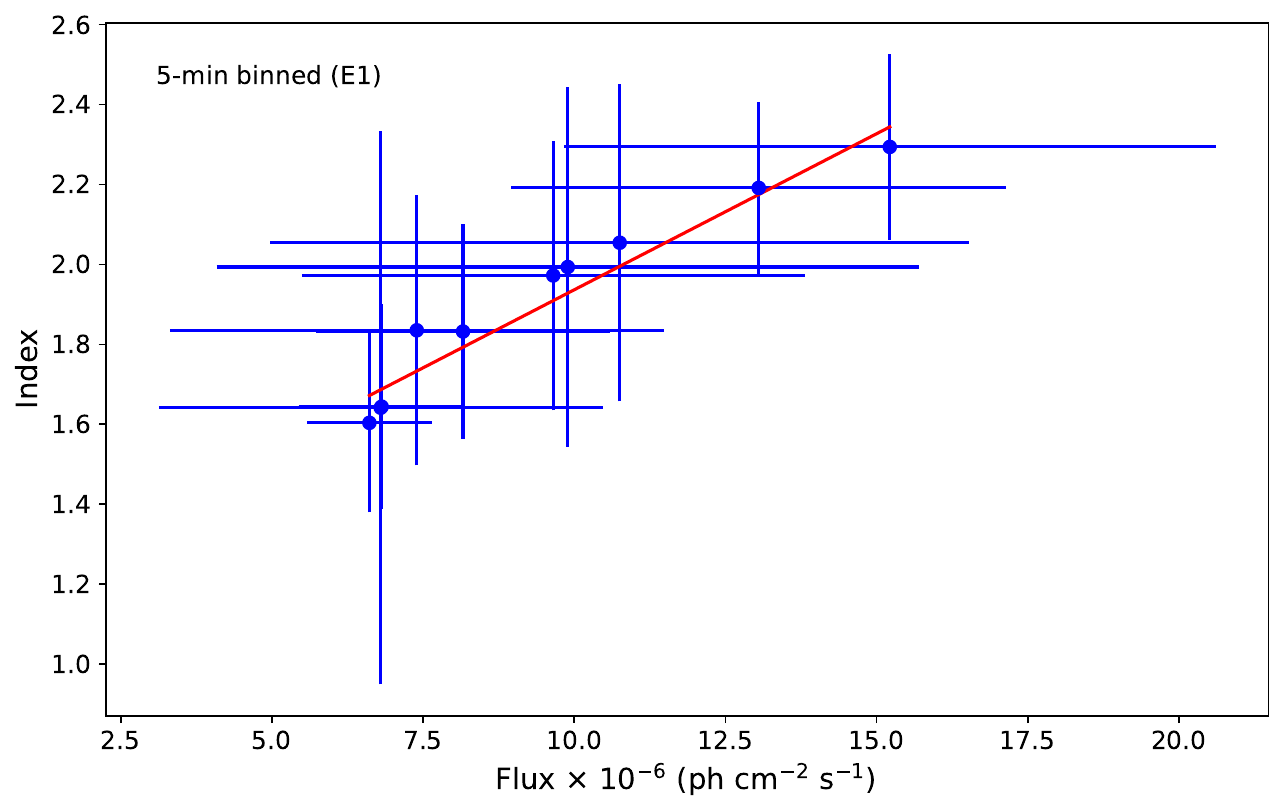}{0.33\textwidth}{(a)}
              \fig{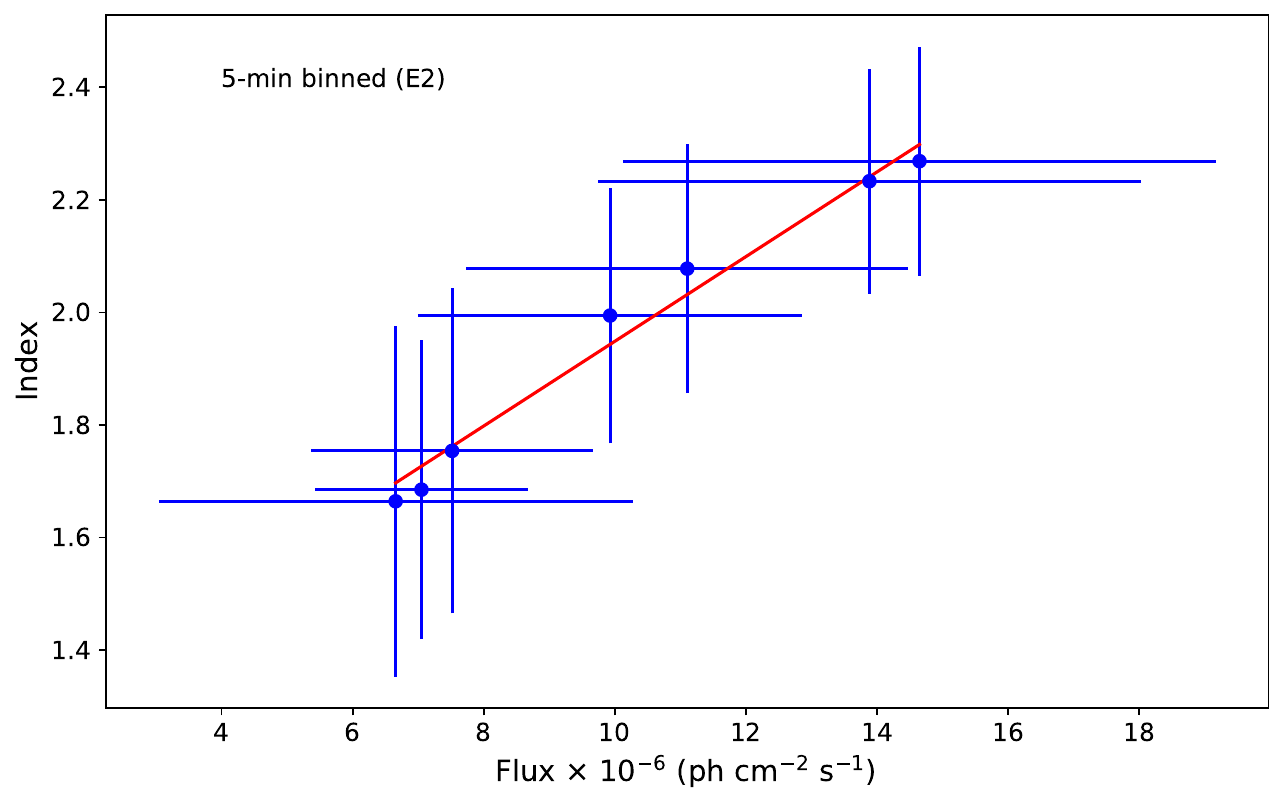}{0.33\textwidth}{(b)}
              \fig{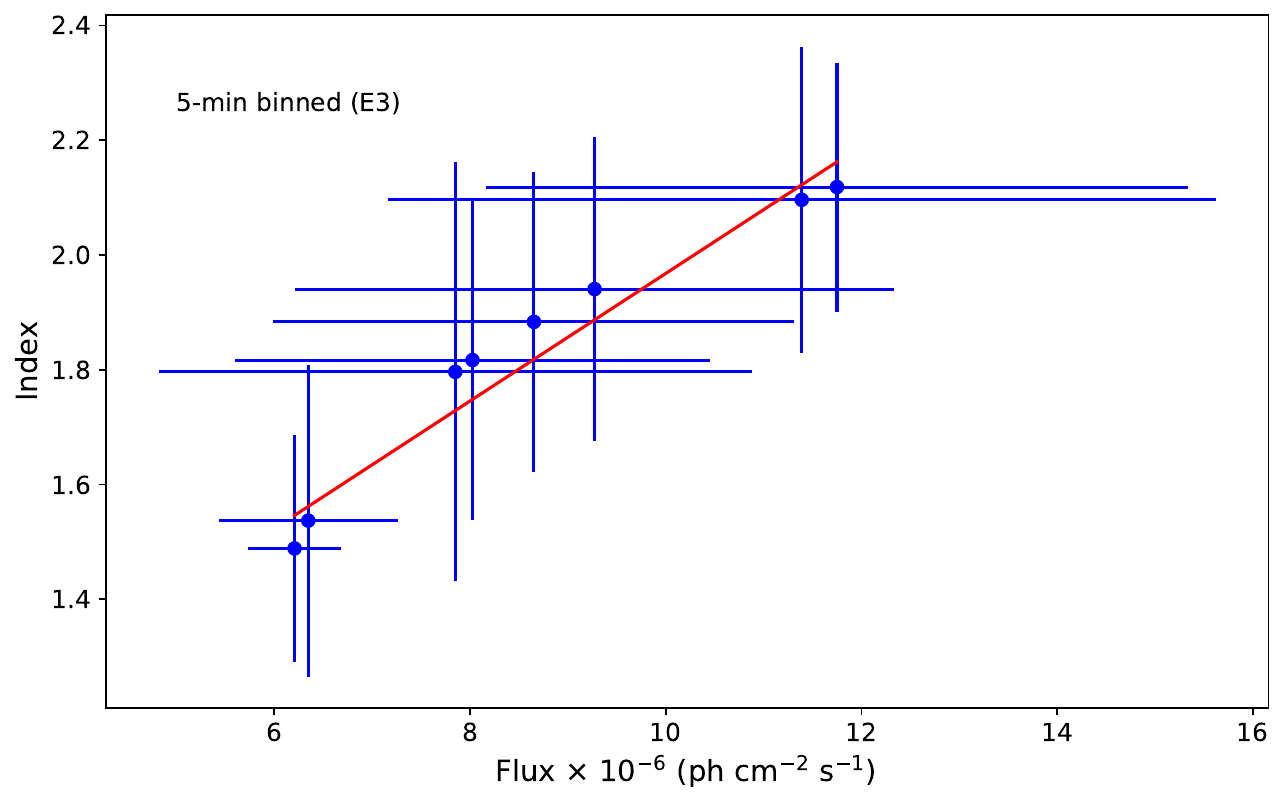}{0.33\textwidth}{(c)}}
    \gridline{\fig{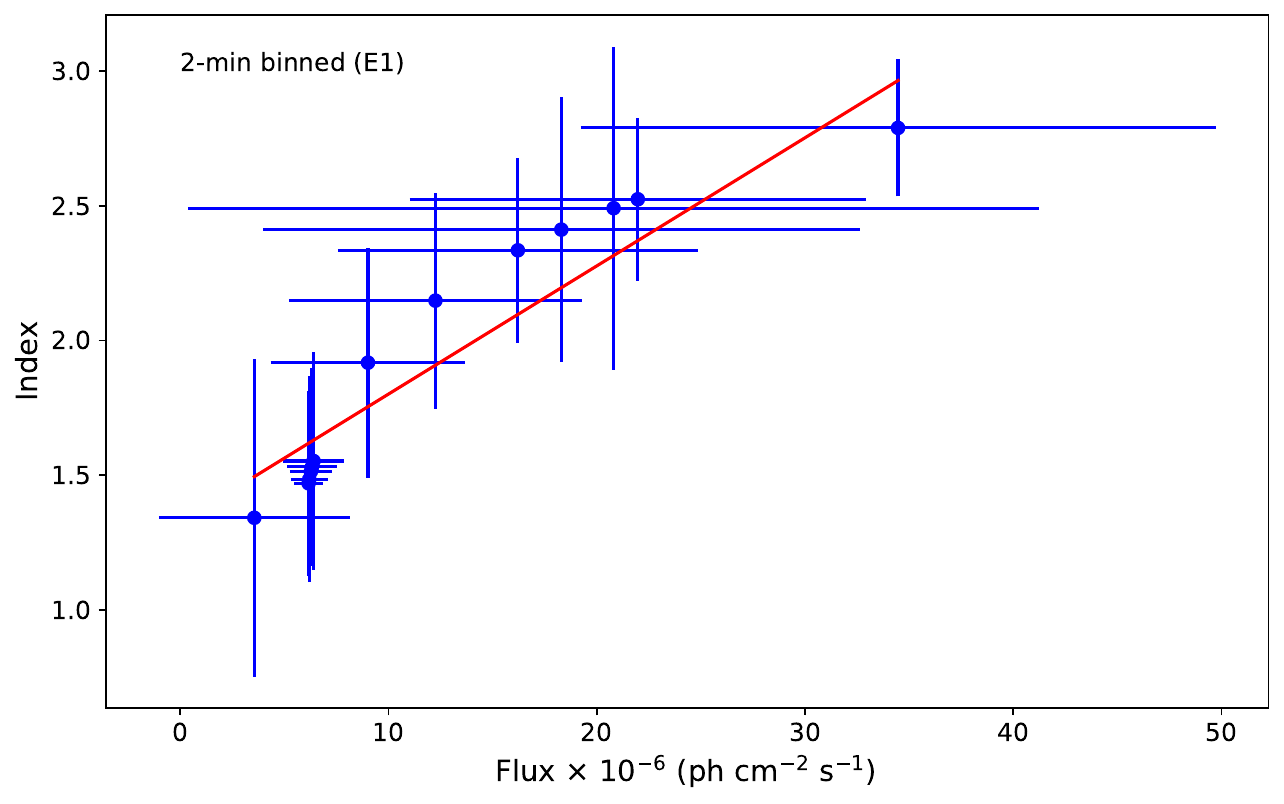}{0.33\textwidth}{(d)}
              \fig{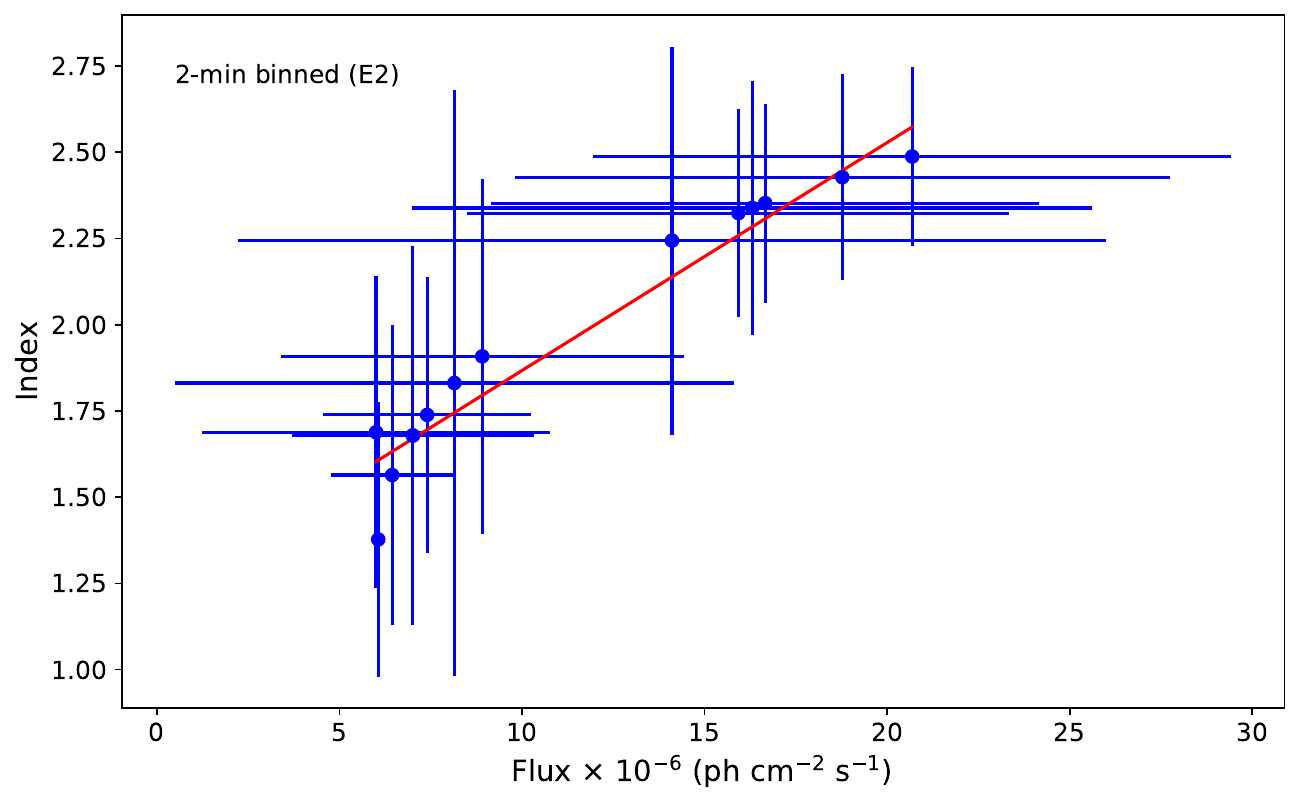}{0.33\textwidth}{(e)}
              \fig{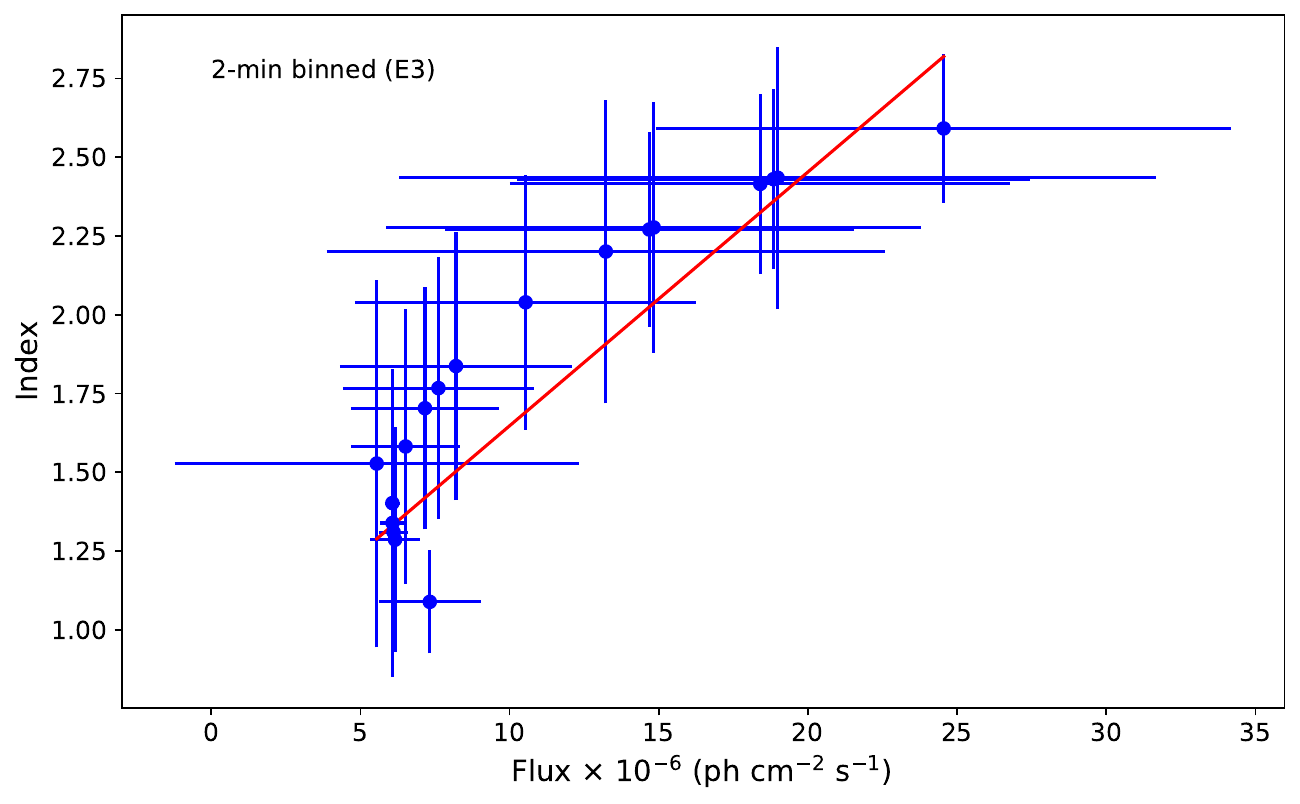}{0.33\textwidth}{(f)}}
    \caption{Variation of PL photon index with $\gamma$-ray flux for the 5-min and 2-min binned light curve of different orbit, mentioned in the figures. The trend of softer-when-brighter is clearly visible.\label{fig:flux_index}}
\end{figure*}

We have investigated the variability of the power law index of BL Lacertae in 1-day binned, 3-hr binned, 5-min binned, and 2-min binned data. We can see that the average PL index is harder than the value (2.2025) mentioned in the 4FGL catalog for each bin size. Table \ref{tab:pearson_corr} shows the average PL index for all the bin sizes. We also investigated the correlation between the $\gamma$-ray flux and PL index using Pearson correlation and the results for each bin size are mentioned in Table \ref{tab:pearson_corr}. No significant correlation between $\gamma$-ray flux and PL index has been found in 1-day, 3-hr, and 95-min (orbital) binned data, but we found a significant positive correlation in 5-min and 2-min binned data as p $>$ 0.90 and r $>$ 0.001. The positive correlations are shown in Figure \ref{fig:flux_index} and they suggest a softer-when-brighter trend for BL Lacertae.

We have generated the gamma-ray SED for each orbit. The log parabola model is used to fit the spectral points in the SED. The $\gamma$-ray SED of E1 is modeled with index $\alpha$ = 1.93 $\pm$ 0.08 and a low curvature $\beta$ $\approx$ 0.04. The $\gamma$-ray SED of E2 is modeled with index $\alpha$ = 1.93 $\pm$ 0.09 and a curvature $\beta$ $\approx$ 0.16 and the SED of E3 is modeled with index $\alpha$ = 1.82 $\pm$ 0.07 and a low curvature $\beta$ $\approx$ 0.04. This shows that orbit E1, E2, and E3 have spectra much closer to a power-law with a very small curvature. We also noticed that above 10 GeV mostly we have upper limits suggesting no high energy photons above 10 GeV are produced.

\subsection{X-ray light curve and spectral variability} \label{subsec:xray_lc_spec}
\begin{figure*}[ht!]
    \gridline{\fig{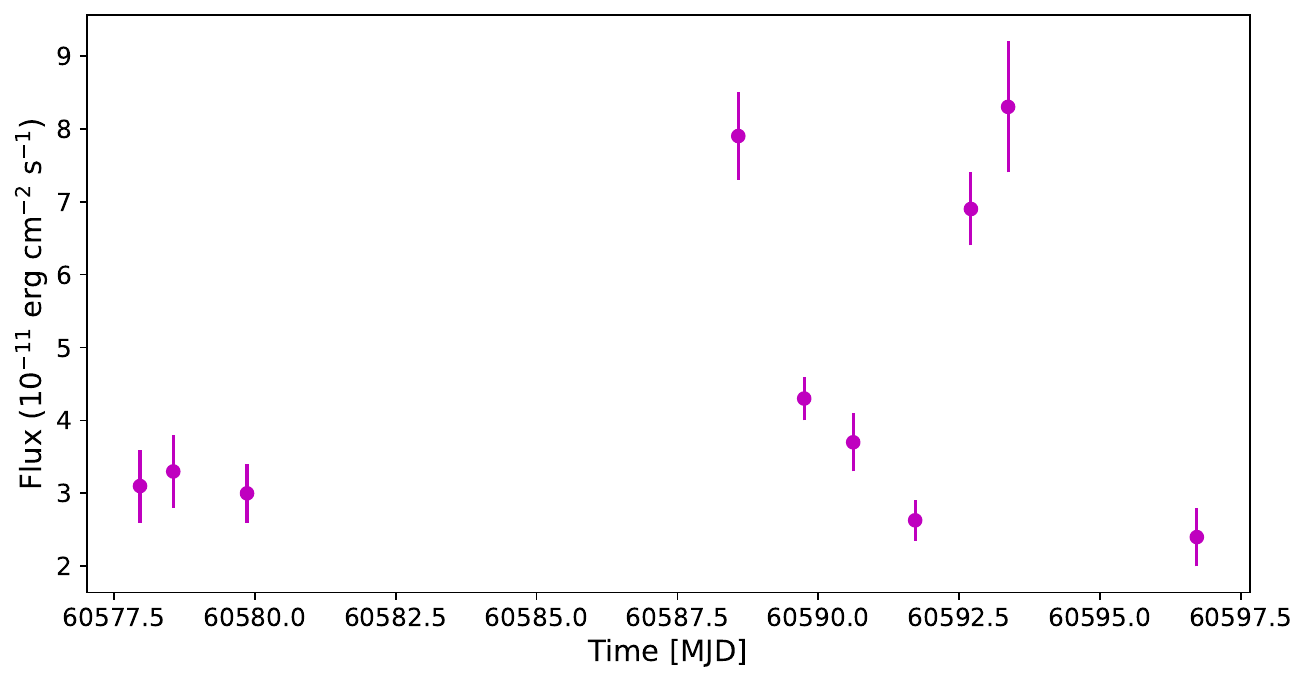}{0.48\textwidth}{(a)}
              \fig{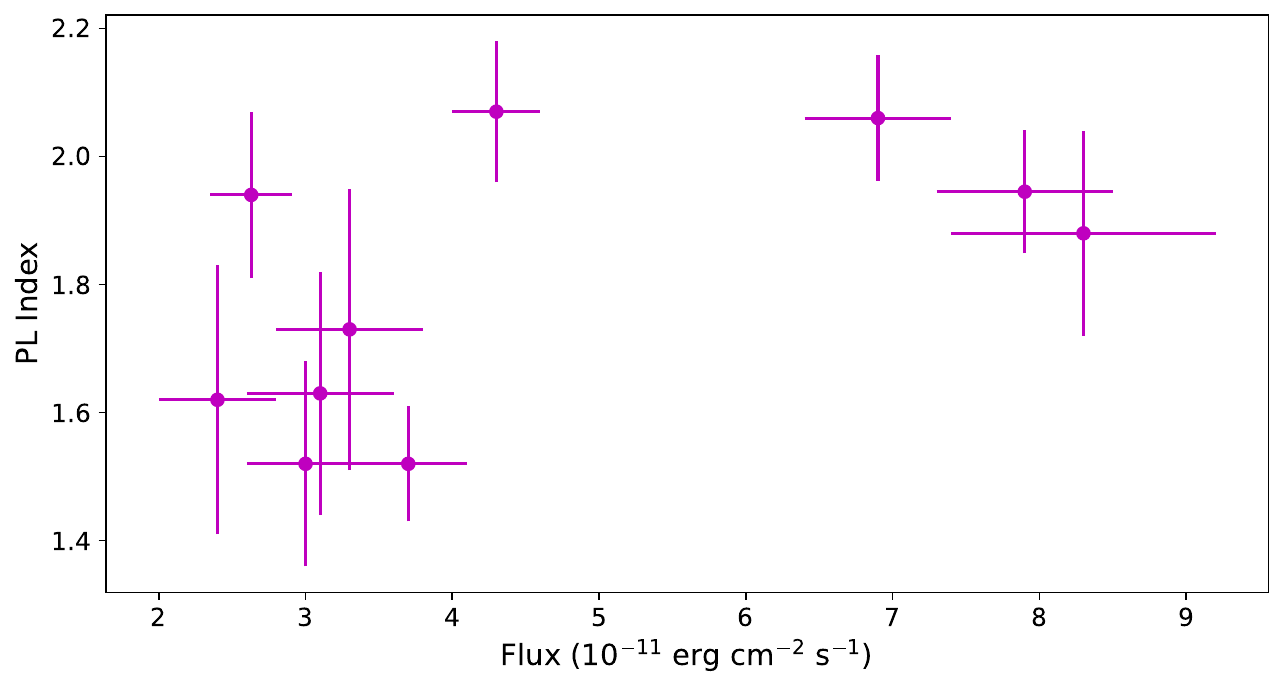}{0.48\textwidth}{(b)}}
\caption{(a) X-ray light curve (b) Flux vs PL index for X-ray observations.
\label{fig:xray_lc_spec}}
\end{figure*}

The X-ray light curve from MJD 60591.72 to MJD 60592.71 is shown in Figure \ref{fig:xray_lc_spec}. With Eqn \ref{eq:flux_doubling_time}, we found the fastest X-ray variability time to be 0.7 days. On MJD 60591, X-ray flux was observed to be (2.63 $\pm$ 0.28) $\times$ 10$^{-11}$ erg cm$^{-2}$ s$^{-1}$ and on next day (MJD 60592) X-ray flux rose to (6.90 $\pm$ 0.50) $\times$ 10$^{-11}$ erg cm$^{-2}$ s$^{-1}$ and even on next day (MJD 60593) the X-ray flux rose to (8.30 $\pm$ 0.90) $\times$ 10$^{-11}$ erg cm$^{-2}$ s$^{-1}$. These consecutive flux enhancements suggest a strong fast X-ray flux variability within a day. Again this suggests a compact emission region involved in the production of such a high variability event.

We have seen a moderate correlation between X-ray flux and PL index with Pearson correlation coefficient r = 0.57 and null-hypothesis probability p = 0.088. As shown in the second panel in Figure \ref{fig:xray_lc_spec}, we can see clearly the softer-when-brighter trend in X-rays like $\gamma$-ray. 

\begin{table}[ht]
    \centering
    \begin{tabular}{l l l l}
        \hline
        Peak & $t_d$ (min) & $T_R$ (min) & $T_D$ (min) \\ \hline
        E1 & 2.13\\
        Peak 1 &  & 4.80 $\pm$ 0.004 & 12.10 $\pm$ 0.002 \\
        Peak 2 &  & 5.07 $\pm$ 0.001 & 6.51 $\pm$ 0.002 \\
        E2 & 4.74\\
        Peak 1 &  & 3.28 $\pm$ 0.001 & 5.16 $\pm$ 0.004 \\
        Peak 2 &  & 3.51 $\pm$ 0.004 & 4.87 $\pm$ 0.001 \\
        E3 & 5.93\\
        Peak 1 &  & 6.06 $\pm$ 0.001 & 7.10 $\pm$ 0.003 \\
        Peak 2 &  & 4.01 $\pm$ 0.002 & 4.29 $\pm$ 0.003 \\ \hline
    \end{tabular}
    \caption{Flux doubling/halving time of each orbit of Fermi-LAT and rise and decay time of each peak in the 5min-binned light curve of BL Lacertae as a result of SOE fitting.}
    \label{tab:rise_decay_table}
\end{table}

\section{Discussion} \label{sec:discussion}
The minute scale flux variability is still a mystery to observational Astronomers in high-energy astrophysics. Many attempts have been made to explain these variabilities \citep{Aleksic_2011, Ackermann_2016, Shukla_2018, 2022A&A...668A.152P}. BL Lacerate object is fourth in the row where minute scale variability is seen and first in the BL Lac types. Other objects where minute scale variability is seen are FSRQs, 3C 279, PKS 1222+216, and CTA 102. In the case of CTA 102, \cite{Shukla_2018} have argued that the minute scale variabilities are produced at a farther distance from the base of the jet due to the dissipation of magnetic islands or protons in a
collimated beam from the base of the jet encountering the turbulent plasma at the end of the magnetic nozzle.

This is not the first time that BL Lacertae has exhibited minute-scale $\gamma$-ray variation, \citealt{2022A&A...668A.152P} had reported a flux halving timescale of $\sim$ (1 $\pm$ 0.3) min. We found that in the current flaring state the BL Lac has surpassed all its previous flux and emerged to be one of the brightest BL Lac-type objects ever detected in Fermi-LAT. This enables us to probe the minute-scale variability in the historically brightest $\gamma$-ray flare ever observed in BL Lacertae. We observed significant minute-scale variation in the 2-min binned light-curve of BL Lacertae where E1 shows 1.49 min, E2 shows 1 min and E3 shows 1.48 min as flux doubling/halving timescale, where E1, E2, and E3 are notations for the three orbits of Fermi spacecraft. The existing $\gamma$-ray emission model is hardly able to explain the minute-scale temporal flux variation.
The central black hole mass of BL Lacertae is 3.8 $\times$ 10$^{8}$ M$_{\odot}$ \citep{2009RAA.....9..168W} and the corresponding event horizon light-cross time is $\sim$ 31 min. Our detected variability time-scale ($\sim$ 1 min) is much shorter than the event horizon light-cross time, suggesting that the enhancement in the $\gamma$-ray originated from a very compact region within the relativistic jet of BL Lacertae. This kind of ultra-fast variability can be explained by three scenarios: the origin of the flare may be a magnetospheric gap with a small volume near the event horizon of the black hole \citep{2007ApJ...671...85N}, magnetic field reconnection in the highly magnetized jet when the emitter moves relativistically in the jet \citep{2005MNRAS.358..113L} and external objects (e.g. stars or clouds) may penetrate in the jet \citep{2010A&A...522A..97A}. In this flaring case, the compact region near the innermost section of the jet is responsible for the GeV band flaring. \citealt{2018ApJ...852..112K} tried to explain the fast very high energy (VHE) variability of the Virgo Cluster radio galaxy M87 (NGC 4486) and the Perseus Cluster galaxy IC 310 (J0316+4119) with the gap-driven magnetospheric $\gamma$-ray emission from the rotating supermassive black holes. The day timescale variability of M87 has been explained by this magnetospheric gap near the event horizon whereas the 5-min timescale variability of IC 310 could not be explained by magnetospheric $\gamma$-ray emission. The minute-scale variability of BL Lacertae with black hole mass M$_{BH}$ $\approx$ 1.7 $\times$ 10$^{8}$ M$_{\odot}$ \citep{2014Natur.510..126Z} is less likely explained by this $\gamma$-ray emission from the magnetospheric gap near the event horizon.

\begin{table}[ht]
    \centering
    \begin{tabular}{c c c c}
         \hline
         Bin size & Avg Index & r & p  \\ \hline
         1-day & 2.11 & -0.39 & 0.006 \\
         3-hr & 2.03 & 0.30 & 4.59 $\times 10^{-6}$ \\
         95-min & 1.94 & 0.39 & 2.25 $\times 10^{-5}$ \\
         5-min (E1) & 1.91 & 0.97 & 5.14 $\times 10^{-6}$ \\
         5-min (E2) & 1.95 & 0.99 & 2.18 $\times 10^{-5}$ \\
         5-min (E3) & 1.83 & 0.97 & 6.07 $\times 10^{-5}$ \\
         2-min (E1) & 1.96 & 0.95 & 1.13 $\times 10^{-6}$ \\
         2-min (E2) & 1.97 & 0.97 & 1.27 $\times 10^{-8}$ \\
         2-min (E3) & 1.86 & 0.92 & 9.83 $\times 10^{-8}$ \\ \hline
    \end{tabular}
    \caption{Average PL index, Pearson correlation coefficient (r), and the null-hypothesis probability (p) for each time bin and orbit as a result of spectral variability.}
    \label{tab:pearson_corr}
\end{table}

We have taken bulk Lorentz factor ($\Gamma$) $\approx$ Doppler factor ($\delta$) \citep{2011ApJ...733L..26A} $\approx$ 16 \citep{2022A&A...668A.152P}. Assuming the emission region to be a spherical blob, we estimated the emission region size using
\begin{equation}
    R \leq c\ t_d \frac{\delta}{1+z},
\end{equation}
where c is the speed to light in vacuum, $t_d$ is the minimum flux doubling/halving time, $\delta$ is the Doppler factor and z is the redshift. Thus, using $\delta$ = 16 and $t_d \sim$ 1 min, we found R $\leq$ 2.68 $\times$ 10$^{13}$ cm. Assuming the jet has a canonical geometry and the distance of the emission region from the central supermassive black hole can be estimated using
\begin{equation}
    R_H \leq 2\ c\ t_d \frac{\Gamma^2}{1+z},
\end{equation}
where $\Gamma$ is the bulk Lorentz factor. Using $\Gamma$ = 16 and $t_d \sim$ 1 min, we found $R_H \leq$ 8.58 $\times$ 10$^{14}$ cm. As we know the size of the broad-line region is around 0.1 pc and hence comparing the distance of the blob with the BLR size suggests that the emission region is much inside the BLR. At this location, the production of minute scale variability is still unknown.

\section{Summary}
In this work, we reported the minute-scale variability in the $\gamma$-ray flare of BL Lacertae in the historically brightest flare. In three different orbits, we have detected temporal flux variability of 1.48, 1, and 1.49 min. We have estimated the upper limit for emission region size and distance of emission region from the central black hole to be 2.68 $\times$ 10$^{13}$ cm and 8.58 $\times$ 10$^{14}$ cm, respectively. We found the softer-when-brighter trend in 5-min and 2-min binned $\gamma$-ray light curve with a highly positive correlation (r $>$ 0.9). We also found fast X-ray variability and X-ray also shows the softer-when-brighter trend along with $\gamma$-ray. 

\bibliography{sample631}{}
\bibliographystyle{aasjournal}

\end{document}